\documentclass[sn-basic]{sn-jnl}

\usepackage{graphicx}%
\usepackage{multirow}%
\usepackage{amsmath,amssymb,amsfonts}%
\usepackage{amsthm}%
\usepackage{mathrsfs}%
\usepackage[title]{appendix}%
\usepackage{xcolor}%
\usepackage{textcomp}%
\usepackage{manyfoot}%
\usepackage{booktabs}%
\usepackage{algorithm}%
\usepackage{algorithmicx}%
\usepackage{algpseudocode}%
\usepackage{listings}%
\usepackage{sidecap}
\newcommand{\tr}{^{\sf T}}

\begin{document}

\title[Category clustering]{\qquad \quad \bf Bi-dendrograms for clustering \\ the categories of a multivariate categorical data set}

\author*[1]{\fnm{Michael} \sur{Greenacre}} \email{michael.greenacre@upf.edu}%

\author[2]{\fnm{Maurizio} \sur{Vichi}} 


\affil*[1]{\orgdiv{Department of Economics and Business, and Barcelona School of Management}, \orgname{Universitat Pompeu Fabra},
\orgaddress{\street{Ramon Trias Fargas 25--27}, \city{Barcelona}, \country{Spain}}}
\affil[2]{\orgdiv{Department of Statistical Science, Università di Roma \textit{La Sapienza}}, \orgaddress{\street{P.le Aldo Moro 5}, \city{Rome}, \country{Italy}}}

\abstract{
The clustering of categories in a multivariate categorical data set is investigated, where the problem separates into that of merging categories of the same variables (i.e., within-variable categories), and combining categories of different variables (i.e., between-variable categories).   
For the within-variable problem, the objective is to arrive at fewer categories (and, consequently, lower data dimensionality) without affecting the essential features of the data set, thereby simplifying the interpretation of any analysis using the categorical variables. 
The categories can be of an ordinal or nominal nature, and this property is respected in the clustering, where only adjacent categories of ordinal variables can be combined.
For the between-variable problem, the objective is to arrive at asmall number of category clusters that typify the observations in the data set.
In this latter problem there is no restriction on which categories can combine, as long as they do not combine within the same variable.
In each of these problems, results are given in the form of a pair of dendrograms stacked one on top of the other, called a bi-dendrogram. 
For the within-variable problem, once all categories within each variable have been merged, the second stage is to cluster the variables themselves.
For the between-variable problem, the second stage is to cluster groups of respondents that fall into the response sets arrived at in the first stage of clustering. 
The approach is illustrated using a sociological survey data set from the International Social Survey Program.
}


\maketitle
\section{Introduction and history of the problem}
We are concerned here with a multivariate data set $\bf X$ ($n\times Q$), of $n$ observations (for example, $n$ respondents in a survey) on $Q$ categorical variables (for example, their responses to $Q$ questions, where a single categorical response is selected for each question). 
Since the context is most often a sample survey, we often use terms such as questions (for variables) and responses (for the observed categories).
The categories can be ordinal or nominal and a missing value category for each variable is also allowed.

For the $q$-th categorical variable, there are $J_q$ possible categorical responses to choose from, so the total number of possible responses is $J = \sum_q J_q$.
For example, in the application treated later, there are $Q=8$ variables and each variable has $J_q = 6$ categories, five substantive categories numbered from 1 to 5 on an ordinal scale of agreement (1 = strongly agree, 2 = somewhat agree, 3 = neither agree nor disagree, 4 = somewhat disagree, 5 = strongly disagree), as well as a missing value category, coded as 9. 
An example of an observation is the vector of eight categorical values  $[\  1\  4\  5\  9\  2\  3\  1\  2\  ]$, where there is a missing value for the fourth variable, otherwise categories from 1 to 5. 
Of interest here is the treatment of the data as categorical, rather than assuming the values 1 to 5 to be on an interval scale.

Since there can be many categories in such a data set, the question arises whether some categories can be combined with minimal loss of information in the data. 
For example, it may be detected that there is little difference in the data set between the categories ``strongly agree" and ``somewhat agree" for a particular variable, and these categories can be combined into one of agreement.
The question is how to identify such similarities between the categories in order to combine them.

This problem goes back to the 1980s, where a certain controversy was generated by two papers of \cite{Carroll:86, Carroll:87} in the \textit{Journal of Marketing Research}.
These authors considered the problem of $Q=2$ categorical variables, with respective numbers of categories $J_1$ and $J_2$, in the context of a correspondence analysis (CA) of a cross-tabulation $\bf N$ ($J_1 \times J_2$) \citep{Benzecri:73, Greenacre:16a}.
It was well-known at the time that in CA distances between categories of the same variable, called \textit{chi-square distances}, were interpretable, but not distances between categories of different variables, since these distances had no specific definition.
The authors Carroll, Green and Schaffer reasoned that the data set could be expressed as an $n\times J$ matrix $\bf Z$ of dummy variables of 0s and 1s, called an \textit{indicator matrix}, where all $J = J_1+J_2$ categories were columns of the same matrix and each row contained two ones indicating the categories selected of the two variables, otherwise zeros.
Thus, they proposed that the distances between all categories, both those between categories of the same variables, as well as those between categories of different variables, could be interpreted in a CA of the indicator matrix $\bf Z$.

\cite{Greenacre:89} published a criticism of this reasoning, again in the \textit{Journal of Marketing Research}, pointing out that the chi-square distances between the dummy variables had no substantive meaning. For example the distances between two categories $j$ and $k$ of the same variable were computed on dummy variables that had no overlap of the ones at all (only one category could be chosen for each variable), and the distance depended only on the marginal frequencies of these two categories.
In a rejoinder that was published in the same journal straight after Greenacre's critique, \cite{Carroll:89} stated that the rejection of their proposal extended to rejecting the grouping of categories in the multivariate case, known as multiple correspondence analysis (MCA).
This rejoinder was unfounded since any grouping of the categories in an MCA solution was not due to their interpoint distances in the indicator matrix, but rather due to their joint bivariate relationships \citep{Greenacre:88a}.

The present article aims to lay this issue to rest by proposing justifiable ways of defining ``proximities of association" between the categories, for the general case of multivariate categorical data, of which the bivariate problem described above is a special case. 
We recognize that the combining of categories of the same variable is a completely different problem from combining categories of different variables, and these need to be considered separately. 
We see no way how they can be put on an equivalent footing, as Carroll, Green and Schaffer attempted to do. 
Thus, there are two parts to this article: combining categories within variables (Section 2), and combining categories between variables (Section 3). 
Each objective has a separate motivation and a separate type of result, with a different interpretation. 
The former within-variable problem aims to merge categories to simplify the coding by reducing the number of categories, which makes subsequent analysis and interpretation easier.
The latter between-variable problem aims to define sets of categories of different variables that characterize the observations in the data set, and then cluster the observations according to such ``response sets".
Section 4 shows how the proposed solutions apply to a data set from a survey in the International Social Survey Program (ISSP). 
Section 5 concludes with a discussion and conclusion.

\section{Combining categories of the same variable (within-variable clustering)}
The multivariate categorical data set on $n$ observations $\bf X$ ($n \times Q$), with the $Q$ variables as columns, can be equivalently coded as an indicator matrix ${\bf Z} = [\ {\bf Z}_1 \ {\bf Z}_2 \cdots \ {\bf Z}_Q \  ]$, with the $J$ categories as columns, and each ${\bf Z}_q$ the $n \times J_q$ submatrix of dummy variables corresponding to the $q$-th variable.
Each row of each ${\bf Z}_q$ contains zeros with a single one indicating the chosen category, so there are exactly $Q$ ones in each row of $\bf Z$. 
The objective is to combine successively pairs of categories within each variable $q$ (i.e., the columns of ${\bf Z}_q$).

This problem was considered by \cite{Greenacre:88b} for bivariate data ($Q=2$) in the form of a two-way contingency table ${\bf N} = {\bf Z}_1\tr{\bf Z}_2$.
The Pearson chi-square statistic $\chi^2$ computed on the contingency table measures the strength of association between the two categorical variables, typically used in the chi-square test of association between the variables. 
Hence, two rows, for example, of the table were merged, by summing their frequencies, if they reduced the chi-square statistic $\chi^2$ the least. 
Since the inertia in CA is equal to $\chi^2 / n$, where $n$ is the row dimension of $\bf X$ and of $\bf Z$ (that is, the sample size which is equal to the grand total of $\bf N$), this criterion is equivalent to minimizing the reduction in the inertia.
Merging is achieved by summing the two chosen rows of $\bf N$, which reduces the row dimension of $\bf N$ by 1.
Equivalently, the two corresponding columns of ${\bf Z}_1$ can be merged by summing and the contingency table ${\bf N}$ recomputed.  
Merging is repeated until all the rows have been aggregated, in which case the inertia is reduced to 0.
The merging of categories of the columns of $\bf N$, equivalently of the columns of ${\bf Z}_2$, can be performed in exactly the same way.
In each case, the sequential mergings of the categories of the two variables can be represented in a dendrogram, where the scale is in terms of the criterion of inertia loss (or, equivalently, chi-square loss). 
The clustering criterion is equivalent to minimizing a weighted  chi-square distance distance between the rows, which is equal in value to the inertia decrease at each step.
This is a weighted form of Ward clustering \citep{Jambu:78, Greenacre:88b}.

\subsection{Clustering the categories}
For more than two categorical variables ($Q > 2$) the approach of \cite{Greenacre:88b} for $Q=2$ is easily generalized. 
One only needs a definition of the inertia for the multivariate case, which already exists in \cite{Greenacre:88a, Greenacre:16a}.
For $Q$ variables there are $\frac{1}{2}Q(Q-1)$ unique bivariate cross-tables ${\bf N}_{qs}$ between variables $q$ and $s$ ($q<s$), each with its respective inertia ${\sf In}_{qs}$. 
The total inertia for the multivariate case, \textsf{InTot}, is defined as the average of these inertia values: 
\begin{equation}
\textsf{InTot} = \sum \sum_{q<s} {\sf In}_{qs} / \left(\frac{1}{2}Q(Q-1)\right)
\label{InTot}    
\end{equation} 
where $\sum\sum_{q<s}$ denotes the double summation over the $\frac{1}{2}Q(Q-1)$ tables. 
To merge two categories, the two corresponding columns of the indicator matrix $\bf Z$ are summed, the cross-tables that are affected by this merging are recomputed as well as the average inertia across all tables. The two categories to be merged are the ones that reduce this average inertia the least.
For $Q=2$ there is only one table and this reduces to exactly the same criterion of \cite{Greenacre:88b}, as explained above.

The number of clustering steps in this algorithm depends on the number of categories for each variable and also whether the scales are ordinal or nominal.
There are three situations at the first step of the algorithm:
\begin{enumerate}
\item 
For ordinal scales, only adjacent categories can be merged, that is, there are $J_q-1$ possible pairs for each variable $q$.
Hence, the number of pairs to consider at the first step of the algorithm is $\sum_q (J_q -1) = J-Q$.
\item 
For nominal scales, any pair can be considered, so at the start there are ${\sum_q} \frac{1}{2}J_q(J_q-1)$ possible pairs. This is also the case when one of the $J_q$ categories is a missing value one.
\item 
When the $J_q$ categories consist of  $J_q-1$ ordinal categories plus an additional missing value category, the ordinal categories can only merge with $J_q-2$ adjacent ones, whereas the missing category can merge with any of the $J_q - 1$ ones.
Hence, there are $\sum_q (J_q-2 + J_q-1) = 2J-3Q$ possible pairs to consider at the first iteration of the algorithm, where $J$ now denotes the total number of categories, including the missing ones.   
\end{enumerate}
The variables can be a mixture of the above three scales, and in all three cases the number of pairs to consider is decreasing at each step.

The merging process ends when all categories of each variable have been merged together, where the average inertia has reduced to zero.
The results can be visualized in a dendrogram where the level of each node corresponds to the value of the average inertia at that step, or alternatively the loss of total inertia since the start.
Thus, in the latter case, the dendrogram scale starts at the original value of the average inertia, before merging starts, and ends at 0.

\subsection{Clustering of the variables}
The clustering of the categories ends with the $Q$ variables internally clustered, and the clustering can continue using the same inertia decomposition, but at the variable level.
In the $Q \times Q$ table of pairwise inertias, the two variables with the largest inertia, which in the sense of variable association are the closest, are clustered first.
The table is updated to a $(Q-1)\times (Q-1)$ table by the averaging method: for example, if questions B and C are clustered, the inertia of question A with the cluster $\{\textrm{B,C}\}$ is the average of two inertias, that of A crossed with B and that of A crossed with C. 
The new average inertia is computed on the reduced table, and the clustering level for the corresponding node is the difference between the inertia before and after clustering. 
In this way, the clustering of the variables also starts with the same average inertia value and descends to 0, as for the categories. 
The result is a dendrogram of the variables, which can be stacked on top of the dendrogram of the categories, with identical scales.
Notice that in the clustering of the variables the algorithm is one of average linkage, whereas the prior merging of categories is one of Ward clustering.

\section{Combining categories of different variables (between-variable clustering)}
The combining of categories from different variables is a completely different problem, with a different objective.
If some categories are grouped together in so-called ``response sets", the idea is that they are associated in some sense in the observations.
To define this sense of association, there are many possibilities. For example, there are $J_1 \times J_2 \times \cdots \times J_Q$ possible combinations of the categories, and the frequencies of these combinations can be counted across the sample to see which is the most frequent.
Then the second most frequent, eliminating from consideration those that contain the categories of the first one, can be sought and so on.
This will necessarily give only min$\{J_q, q=1,\ldots,Q\}$ combinations in the results, so an alternative algorithm is required, preferably resulting in an agglomerative hierarchical clustering.

\subsection{Clustering the categories}
In simple CA the idea of association between two categories of the row and column variable, is encapsulated in the standardized residuals of the relative frequencies with respect to the row and column margins. 
Suppose that ${\bf P} = {\bf N} / n$ is the correspondence matrix corresponding to the contingency table $\bf N$ with grand total $n$. 
$\bf P$ has grand total 1 and can be considered the observed discrete bivariate distribution of the two categorical variables. 
Let ${\bf r} = {\bf P 1}$ and ${\bf c} = {\bf P}\tr {\bf 1}$ be the vectors of row and column sums of $\bf P$ (i.e., the marginal distributions), where the vectors of ones $\bf 1$ are of appropriate dimensions.
The matrix of standardized residuals is:
\begin{equation}
    {\bf T} = {\bf D}^{-1/2}_r ({\bf P} - {\bf r}{\bf c}\tr) {\bf D}^{-1/2}_c  
    \label{std_residuals}
\end{equation}
with elements $t_{ij} = (p_{ij}-r_ic_j) / \sqrt{r_ic_j}$.
The sum of squares of $\bf T$ is equal to the total inertia of the contingency table, and is equal to $\chi^2 / n$, as already pointed out in Section 2.
The contribution $(p_{ij}-r_ic_j)^2 / (r_ic_j)$ of each $(i,j)$ combination to the inertia can be used to measure the strength of association of row category $i$ and column category $j$.
In fact, the positive standardized residuals in the matrix (\ref{std_residuals}) are the ones indicating positive association, and the higher the residual, the stronger the association.

The above reasoning can be extended to the multiway case by considering all pairwise contingency tables ${\bf N}_{qs}$ of the $Q$ categorical variables.
Suppose that ${\bf T}_{qs}$ is the matrix of standardized residuals for table ${\bf N}_{qs}$.
Let $\bf T$ now denote the $J \times J$ block matrix of all the matrices ${\bf T}_{qs}$ (for $q\neq s$), including on the diagonal blocks the matrices ${\bf T}_{qq}$ corresponding to (\ref{std_residuals}) for the cross-tabulations of each variable with itself.
Notice that the sum or squared values in the half-triangle of matrices ${\bf T}_{qs}$ (for $q < s$), that is, the sum of the inertias of all unique pairwise cross-tables, divided by $\frac{1}{2}Q(Q-1)$ gives the measure of total inertia, \textsf{InTot}, defined in (\ref{InTot}).

It is on the matrix $\bf T$ that agglomerative hierarchical clustering is performed, for example using complete or average linkage.
This is a matrix of similarities and need to be reversed in value if a dissimilarity-based algorithm is used.
The scale of the dendrogram is in terms of the inertia contributions (the squares of the similarities), going from high at the initial stage to low. 
To be directly related to the total inertia \textsf{InTot}, these squared values should be divided by $\frac{1}{2}Q(Q-1)$.
Hence, once again, the scale of the clustering is in terms of loss of total inertia, but this time all the inertia is not lost at the end of the process.
Notice that the off-diagonal values of each ${\bf T}_{qq}$ are highly negative, leading to high dissimilarities, which precludes categories of the same variable to be chosen by the algorithm.

From the dendrogram, a set of category clusters is deduced, say $G$ clusters.
The number of iterations of the algorithm until all the chosen clusters are formed is equal to $J-G$.
The clusters define subsets of responses, which can be interpreted and the respondents can each be assigned to a cluster, as described in the following subsection.
Once all respondents are assigned to the clusters, their frequencies across all the categories can be used in a further clustering.  

\subsection{Clustering the category clusters}
The above algorithm terminates when all the categories have been clustered, forming a set of $G$ clusters formed, each of which  contain a subset of categories of different variables.
The whole sample can now be allocated to one or more of these clusters by a process that is reminiscent of the first step of a k-means algorithm: that is, assign each observation to the cluster to which it is the ``closest".
Here we choose a measure of closeness using the concept of \emph{weighted resistance}, which takes the ordinality of the scales into account.
This concept is the basis of an alternative algorithm for between-variable category clustering, described in the Supplementary Material.
Interestingly, the concept of resistance is due to the 19th-century mathematician Charles Dodgson, alias Lewis Carroll, the author of `Alice in Wonderland' -- see \cite{Dodgson:1876}.

For example, suppose that one cluster is defined by the combination of categories A1, B5 and C5 for the first three variables A, B and C.
The weighted resistance of an observation that has values 2, 3 and 4 for these three respective variables would be $1+2+1=4$, i.e. changing 1 position on variable A, 2 on B and 1 on C.
This sum of absolute differences should be averaged so that this value can be compared to weighted resistances to other clusters that are defined by more or less than three categorie.
In this example, the average weighted resistance of the observation to be assigned to this cluster is $4/3 = 1.333$. 
The same observation's average weighted resistances to the other clusters are computed and the observation is then assigned to the cluster with respect to which it has the lowest resistance for being assigned. 
It can happen that there is a tie for lowest value, in which case the observation is assigned ``fuzzily" in fractions to the clusters.
For example, if there is a tie for an observation to be assigned to clusters 1 and 2, then the observation is assigned 0.5 to each cluster.

After every observation has been assigned to a cluster, either crisply or fuzzily, a $G \times J$ cross-tabulation can be made of the $G$ clusters by the complete set of $J$ categories.
The chi-square distance from correspondence analysis is a convenient distance function between the $G$ clusters, and again a hierarchical clustering can be made of the clusters, using the same method as before, for example average linkage. 

In the Appendix, an alternative algorithm for between-variable  category clustering is explained based on the weighted resistance measure in the first stage of the clustering.

\section{The `women working' data set}
Our methods are applied to a data set from the International Social Survey Program (ISSP), specifically its fourth survey in 2002 on Family and Changing Gender Roles \citep{ISSP:02}.
The Spanish sample ($N=2421$) was chosen, of which 314 respondents had some missing values, so a reduced data set of $N=2107$ respondents is also considered to include respondents with no missing data.

The respondents were asked to express their agreement/disagreement to eight statements related to working women: 

\begin{description}
\item[\textbf{A.}]
A working mother can establish a warm relationship with her child
\item[\textbf{B.}]
A pre-school child suffers if his or her mother works
\item[\textbf{C.}]
When a woman works, family life suffers
\item[\textbf{D.}]
What women really want is a home and kids
\item[\textbf{E.}]
Running a household is just as satisfying as a paid job
\item[\textbf{F.}] 
work is best for a woman's independence
\item[\textbf{G.}] 
a man's job is to work; a woman's job is the household
\item[\textbf{H.}]
Working women should get paid maternity leave
\end{description}

\noindent
Respondents had to choose an option from the following five-point scale:
1 -- strongly agree, 2 -- somewhat agree, 3 -- neither agree nor disagree, 4 -- somehat disagree, 5 -- strongly disagree, as well as an option of ``Don't know", which is treated as a missing value category.

This data set has been analysed by multiple correspondence analysis in chapters 9 and 10 of \cite{Greenacre:10b}.

\begin{SCfigure*}
 \includegraphics[width=7.6cm]{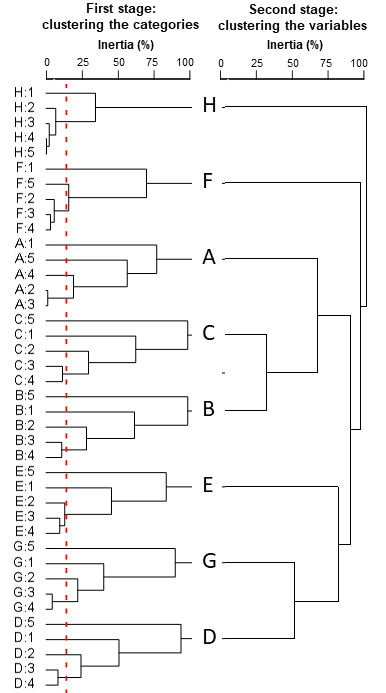}   
    \caption{Bi-dendrogram clustering of the ordinal categories of the same variables, followed by clustering of the variables. In both cases, the clustering criterion is the minimum reduction of inertia, expressed here as a percentage of the total inertia ($=0.2215$) and accumulated across the clustering. The letters refer to the variables, and the numbers the response categories, listed in Section 4.}
    \label{within_clustering}
\end{SCfigure*}

\section{Results}
Figure \ref{within_clustering} shows the result for the clustering of the categories within variables for the reduced data set (i.e., without the respondents with missing data) until the variables are all internally clustered, followed by the clustering of the variables themselves.
The horizontal dashed line shows the cutpoint for significant clustering, based on the result for two-way tables by \cite{Greenacre:88a} (see the table of critical values in Appendix A, Table A.1, of \cite{Greenacre:16a}).
Using this threshold as a guideline, it is generally only the extreme categories 1 and 5 that remain separated, with the exception of variable H. 
B2 and C2 (agree to children and family suffering if the mother works) also separate significantly from the rest. Otherwise all the `moderate' categories 2, 3 and 4 are merged.
The categories that cluster the last, H1,F1,A1,C5,B5,E5,G5 and D5 all correspond to the extreme categories in favour of women working. 
As for the clustering of the variables, it is no surprise that B and C cluster first, since their wording is very similar, and the patterns of their merging categories are almost identical. 
Similarly, the next pair of variables to cluster is G and D, both relating women to their home.
The last to cluster is variable H, which has low association with all the variables, being a separate issue not directly related to the woman's role at home or at work.

\begin{SCfigure*}

 \includegraphics[width=8.5cm] {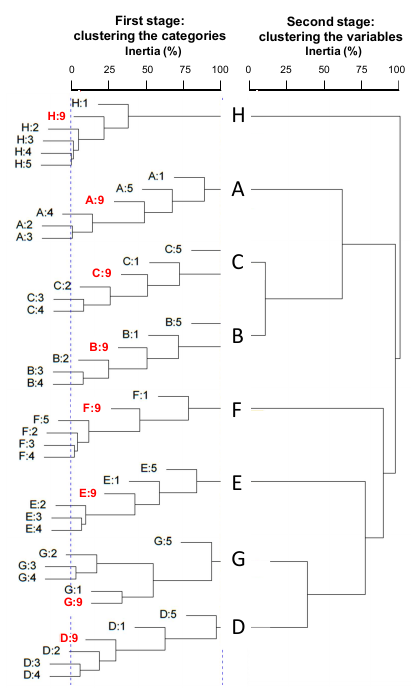}  
    \caption{Bi-dendrogram lustering of the categories, including the missing categories (coded as category 9), within each of the variables, followed by clustering of the variables. In both cases, the clustering criterion is the minimum reduction of inertia, expressed here as a percentage of the total inertia ($=0.2813$) and accumulated across the clustering The first stage dendrogram is of the hanging form, where zero is indicated by the blue dotted line.}
    \label{within_clustering_missing}
\end{SCfigure*}

When the missing value categories are included in the complete data set, Figure \ref{within_clustering_missing} shows that they generally merge with other categories at a high level, but not as high as the highest categories in favour of women working already seen in Figure \ref{within_clustering}. 
Notice the increase in total inertia when the missing value categories are included, going from 0.2215 in Figure \ref{within_clustering} to 0.2813 in Figure \ref{within_clustering_missing}.

\begin{figure}[h]
 \begin{center}
 \includegraphics[width=13.5cm]{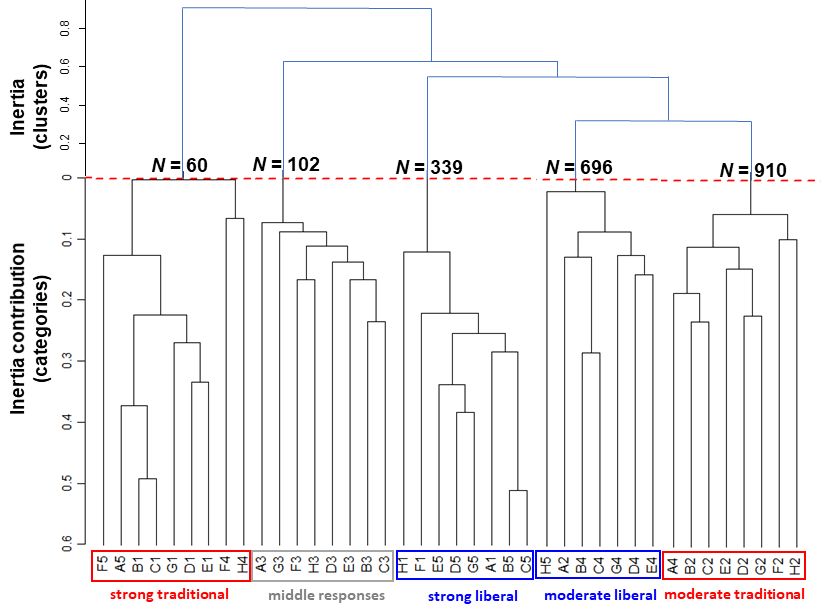}
    \caption{Bi-dendrogram clustering of the categories between variables, using average linkage on the standardized residuals in the first stage, and then average linkage on the chi-square distances between the clusters in the second stage.}
    \label{between_clustering_standardized}
  \end{center}  
\end{figure}

The clustering of the categories between-variables is shown in Figure \ref{between_clustering_standardized}, in this case only for respondents with no missing data.
The categories 2 and 4 are referred to as ``moderate" and the categories 1 and 5 as ``extreme". 
The adjectives ``traditional" and ``liberal" are used depending on the wording of the questions.
Here we see the groupings into strong traditional, moderate traditional, strong liberal and moderate liberal attitudes to women working, as well as all the missing categories grouping together. 
These five clusters have sample sizes as indicated on the dendrogram. 
The largest group is the moderate traditional one (910 respondents, 43\% of the Spanish sample), tending to prefer working women to stay at home and look after the house and children, whereas 696 (33\%) are moderate liberal, tending to the opposite attitude. The fact that these two moderate groups join first in the clustering concords with the main contrast in these data, which opposes moderate views versus extreme views, irrespective of the direction of the attitude. 
There are more strongly liberal than strongly traditional (339, 16\%, compared to 60, 3\%, respectively) and overall more liberal (1035, 49\%) than traditional (970, 46\%). 
The remaining 102 (5\%) are characterized by the tendency to give the middle responses, ``neither agree nor disagree".

\section{Discussion}
Notice once again how different the objectives and results are of the within-variable and between-variable category clusterings, which precludes any attempt at clustering them all as one set (cf. the Carroll-Green-Schaffer papers described in Section 1).
The within-variable clustering aims to simplify the coding by merging categories that are similar in the sense that separately they carry little additional information in the data set.
The between-variable clustering aims to arrive at response sets of combinations of categories from different variables, which then identify subsets of respondents with similar responses.
This latter goal can similarly be achieved by a clustering of the respondents using a suitable between-respondent distance measure (e.g., the distances based on the major dimensions of a multiple correspondence analysis \citep{Greenacre:16a} and then investigating \textit{a posteriori} the response characteristics of the emergent clusters.   

Both forms of category clustering, within-variable and between-variable, are an alternative form of dimension-reduction and simplify the interpretation and understanding of the multivariate categorical data.    
The two forms can be applied serially, by first reducing the number of categories by the within-variable clustering, followed by the between-variable solution.
The final count of observations classified into each cluster in the second phase (as in Figure \ref{between_clustering_standardized}) is a useful final summary of the results.

\clearpage


\section*{Appendix}
\medskip
\leftline{\bf Alternative between-variable clustering of categories based on weighted resistance}

\medskip

In Section 3.1 we have used the standardized residuals to cluster categories between-variables. Having obtaioned clusters of categories, respondents were assigned to clusters based on minimum weighted resistance. 
An alternative idea for clustering categories between variables is based on the same idea of weighted resistance.

Suppose that the cross-tabulation of the $I$ respondents between columns $j$ and $k$, which are categories of different variables, is given by Table \ref{Cross-tab}.

\begin{SCtable}[1.5][ht]
    \large
    \centering
    \begin{tabular}{ccc}
        \ \quad & \multicolumn{2}{c}{$k$} \\
        $j$ \quad & 0 & 1\\
        \hline
        0 \quad & $a$ & $b$\\
        1 \quad & $c$ & $d$\\
        \hline
    \end{tabular}
    \caption{Cross-table of two categories $j$ and $k$ of different variables. A 1 indicates that the corresponding category is chosen, a zero not. }
    \label{Cross-tab}
\end{SCtable}

\noindent
The dissimilarity $(b+c)/I$ is the proportion of the sample that needs to be persuaded to change opinion so that all $I$ observations give the same response $j$ and $k$ to the respective questions.
Let us call this index of dissimilarity the \textit{resistance}, a term borrowed from the field of collective choice and attributed to Charles Dodgson (alias Lewis Carroll, the author of Alice in Wonderland) \citep{Dodgson:1876}.  

A variation of this index, for ordinal data, is to introduce a weighting into this ``change of opinion" resistance measure. 
This will depend on what scale values are taken by categories $j$ and $k$. 
For example, if a person is being asked to change a response from 2 to 3, say, on the ordinal scale, then that can be counted as 1 unit of resistance, whereas changing from 2 to 5 is 3 units of resistance. These units of resistance can be counted across the sample who did not answer $k$ to the second question and similarly for those who did not answer $j$ to the first, to arrive at a \textit{weighted resistance}. 
This count should still be divided by $I$ to normalize with respect to the sample size, although this does not give a well-defined proportion.

The following is an example to illustrate the idea. 
Suppose the $2\times 2$ table tabulating two responses to questions B and C are given in Table \ref{Cross-tab2} (this is an actual table from the data to be used in the application in Section 3).

The number of respondents, out of the total sample of 2107,who need to be persuaded to change their response to 5 for both questions is $63 + 35 = 98$, giving a resistance of $98/2107 = 0.0465$, that is, 4.65\% of the sample.
However, the complete cross-tabulation of questions B and C, given in Table \ref{Cross-tab3}, shows the breakdown of those respondents whose opinion needs to be changed, for example for changing to category C5 there are 13 who answered C1, 17 who answered C2, etc., and for changing to B5, there are 3 who answered B1, 12 who answered B2, etc.

\begin{SCtable}[2][ht]
    \centering
    \begin{tabular}{crr}
        \ \quad & \multicolumn{2}{c}{C5} \\
        B5 \quad & 0 & 1\\
        \hline
        0 \quad & 1947 & 35\\
        1 \quad & 63 & 62\\
        \hline
    \end{tabular} 
    \caption{Cross-table of two categories B5 and C5, i.e. category 5 of variable B and category 5 of variable C. For example, 62 respondents gave responses 5 to both variables, whereas 62 gave 5 for B but another response to C.}
    \label{Cross-tab2}
\end{SCtable}

\begin{SCtable}[2][ht]
    \centering
    \begin{tabular}{crrrrr}
   \ \quad &  C1 & C2 & C3 & C4 & C5 \\
   \hline
B1\quad &  80 &  34 &   4 &   3 &   3 \\
B2\quad &  52 & 673 &  65 & 154 &  12 \\
B3\quad &  12 &  82 &  82 &  59 &   2 \\
B4\quad &   8 & 182 &  79 & 378 &  18 \\
B5\quad &  13 &  17 &   7 &  26 &  62 \\
  \hline
    \end{tabular}
    \caption{Cross-table of all response categories categories of variables B and C. The four counts in the last row: $13+17+7+26 = 63$, the count in the last row of Table \ref{Cross-tab2}. That is, Table \ref{Cross-tab2} is the aggregations of the first 4 rows and first 4 columns of this table, labelled 0 (did not give response 5).}
    \label{Cross-tab3}
\end{SCtable}

\noindent
The weighted sums are thus equal to $4\times 13 + 3\times 17 + 2\times 7 + 1\times 26 = 143$ (for those answering C1 to C4 changing to C5) and $4\times 3 + 3\times 12 + 2\times 2 + 1\times 18 = 70$ (for those answering B1 to B4 changing to B5), giving a total of $143+70=213$.
This total is again normalized with respect to the sample size $I=2107$: $213/2107 = 0.101$.
These values order the pairs in terms of their weighted resistances.

To give a general formula, suppose the $j$-th and $k$-th categories of two different variables are being evaluated, and suppose that the crosstabulation of these two variables is denoted by $\bf N$, where $j$ is the $j$-row and $k$ the $k$-th row of $\bf N$.
The total sample size $I$ is the sum total of the values in $\bf N$.
Then the weighted resistance is:
\begin{equation}
    \left(\sum_u |u-k| \, n_{ju} + \sum_v |v-j| \, n_{vk} \right) / I
\end{equation}
where the summations are over the columns and rows respectively of $\bf N$.

The above computation is made for every pair of categories from different variables, and the pair that is chosen to be clustered, is the one with the least weighted resistance. 
Clustering continues using the complete linkage algorithm, so that at any particular level of clustering, clusters formed at that level contain categories that have at most that level of weighted resistance pairwise.

\begin{figure}[ht]
 \begin{center}
 \includegraphics[width=13cm]{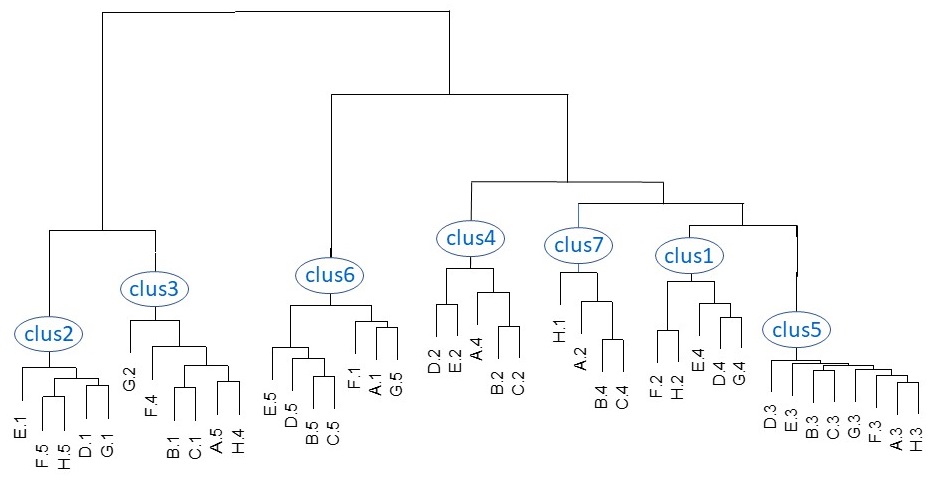}
    \caption{Clustering of the categories between variables, followed by the clustering of the clusters themselves. Both cluster analyses use complete linkage.}
    \label{between_clustering}
  \end{center}  
\end{figure}

As before, clustering of categories continues until all have been clustered, and in this case seven clusters are formed, shown in Figure \ref{between_clustering}.
Respondents are assigned to the clusters using the same weighted resistance measure, as described in Section 3.2.
On the left, clusters 2 and 3 represent 4.0\% of the total Spanish sample who are associated with traditional attitudes, strongly opposed to women working. Cluster 3 (2.6\% of the sample) is especially opposed on child and family issues, and cluster 2 (1.4\%) even more radically anti-feminist. Cluster 4 (30.8\%) is moderately traditional on child and family issues. 
Cluster 5 (2.8\%) brings all the middle responses together.
The remaining three clusters, with a majority of 62.3\% in the sample, all reflect liberal attitudes in favour of women working.
On the moderate side are clusters 1 and 7, with cluster 1 (27.5\%) liberal in favour of women working outside the home and cluster 7 (26.4\%) liberal on the effect on children and the family. Finally, cluster 6 (8.4\%) is strongly liberal in favour of women working in generally, which one might call the pro-feminist group. 

Based on the $G \times J$ cross-tabulation of clusters by respondents, in this case $7 \times 40$, the clusters can themselves be clustered, as before, using the chi-square distance from correspondence analysis ibetween clusters.

\begin{figure}[ht]
 \begin{center}
 \includegraphics[width=13cm]{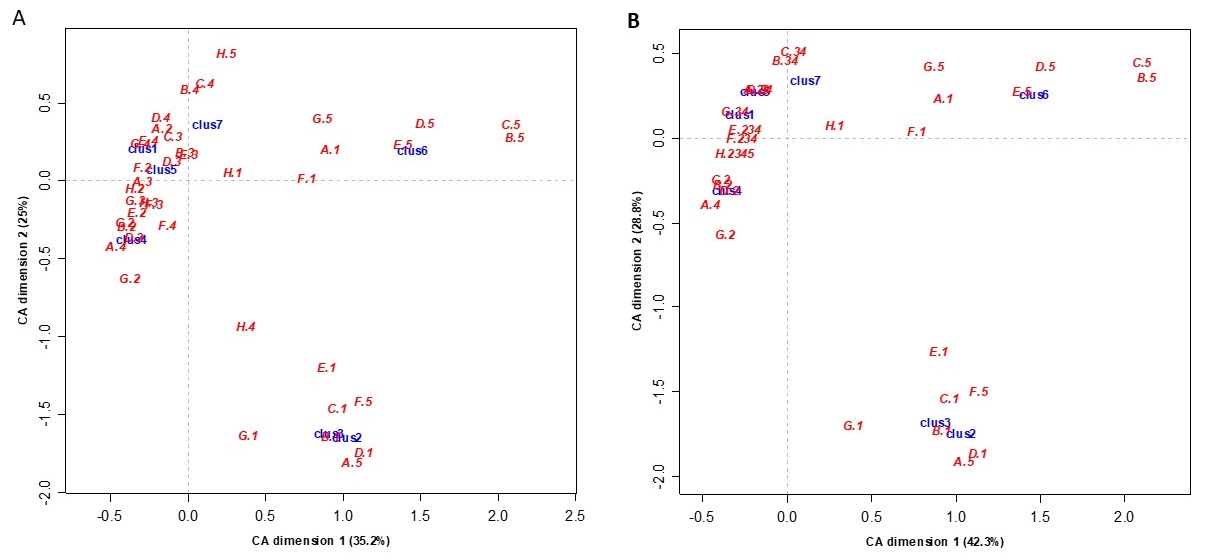}
    \caption{Correspondence analysis of the clusters cross-tabulated with the response categories. A. Using all 40 categories. B. Using the merged categories from the within-variable clustering.}
    \label{between_CA}
  \end{center}  
\end{figure}

The clustering of the clusters is further supported by performing the CA of the $7\times 40$ frequency table of the clusters by variable categories. 
Figure \ref{between_CA}A shows the result for all 40 categories, whereas Figure \ref{between_CA}B simplifies the result by merging categories according to the within-variable clustering in Figure \ref{within_clustering}.
In this way the two clusterings of categories, first within-variables and then between-variables, are combined. 
In both maps of Figure \ref{between_CA}, the proximities of the clusters generally agree with their clustering in Figure \ref{between_clustering}. 

\bibliography{CatClust}
\end{document}